# Large-Scale Cost-Effective Mid-Infrared Resonant Silicon Microstructures for Surface-Enhanced Infrared Absorption Spectroscopy

Pooja Sudha[1], Anil kumar[1], Kunal Dhankar[2], Khalid Ansari[1], Sugata Hazra[2,3], Arup Samanta[1,3]

[1]Department of Physics, Indian Institute of Technology Roorkee, Roorkee, Uttarakhand 247667, India
[2]Department of Biotechnology, Indian Institute of Technology Roorkee, Roorkee, Uttarakhand 247667, India
[3]Centre for Nanotechnology, Indian Institute of Technology Roorkee, Roorkee, Uttarakhand 247667, India.

The mid-infrared (MIR) region spanning from, $\lambda$ = 2.5 µm – 25 µm, is crucial for elucidating the unique biochemical signatures of microorganisms. The MIR resonant structures turned out to facilitate exceptional performance owing to the enhance electric field confinement in the nano-sized aperture. However, the extension of such technique in bacteria-sensing remains limited, primarily due to its micrometre size. This work is the first demonstration of a MIR resonant structure, the gold-coated micro-structured inverted pyramid array of silicon exhibiting light-trapping capabilities, for the bacteria detection in entire MIR range. The electric-field localization within the micro-sized cavity of inverted pyramid amplifies the light-matter interaction by harnessing surface plasmon polaritons, leading to improved detection sensitivity. The confinement of electric field is further corroborated by electric-field simulations based on finite element method. In particular, we observed notable enhancement in both the quantitative and qualitative detection of *Escherichia coli* (E. coli) and *Staphylococcus aureus* (S. aureus) for the bacteria cell concentration of $4 \times 10^5$ cells/ µl, reflecting the efficacy of our detection method. Furthermore, the cost-effective micro structured silicon is fabricated using metal-assisted chemical etching method with the lithography-free method, along with the capabilities of wafer-scale fabrication. Moreover, our device configuration even demonstrates the characteristics of reusability and reproducibility offers substantial benefits over conventional detection schemes. Consequently, this CMOS technology-compatible biosensor signifies promising ways for the integration of this technology with forthcoming bio-applications.

**Keywords:** Surface-Enhanced Infrared Absorption (SEIRA), Localized Surface Plasmon Resonance, Label-free detection, Silicon-microstructures,

## 1. Introduction

Infrared (IR) spectroscopy exhibits prominence as a non-invasive analytical technique that enable the detailed and characteristic exploration of molecules notably in MIR range [1]. Along with its non-destructive nature, IR spectroscopy offers label-free detection, which enhances its applicability for microorganism detection. In recent years, the development of advanced bio-sensing technologies has revolutionized various fields including healthcare, environmental monitoring, and food safety [2, 3]. However, the discrepancy among the size of micro-organisms and detection wavelength limits the sensitivity of the IR signals. To enhance these signals and the overall performance of biosensors, the integration of light-trapping materials has emerged as a promising candidate. Light-trapping materials, characterized by their ability to confine light within nano-structured architectures, offer unparalleled opportunities for signal amplification and improved detection limits in bio-sensing applications [4-6]. These materials exploit principles of photonics, plasmonic, and metamaterials to manipulate the behaviour of light at subwavelength scales, thereby enhancing interaction with analytes of interest. Following this, several antenna structures like hole antennas [7], bow-tie antennas [8], dipole antennas [9] and metamaterials like split-ring resonators [10], nanodiscs [11, 12], triangles [13], and cross-shaped [14], which resonate at the MIR range are employed for enhancing the infrared signals. These enhanced signals from aforementioned surfaces are then termed surface-enhanced infrared absorption signals and this spectroscopy is known as surface-enhanced infrared absorption spectroscopy (SEIRA). However, these structures are used mainly for electric field confinement in the nanometre range, but they are not efficiently suitable for bacteria detection as the bacteria size extended from ~ 1 µm – 5 µm. These antenna configurations are generally characterized by a limited frequency range and even in instances, where they offer a broad bandwidth, they are inadequate in providing comprehensive spectrum data. Besides, the most frequently used microorganism-sensing materials consist of metal nanoparticles mainly gold and silver [14, 16]. These particles do not exhibit resonance in the MIR range, leading to limited signal enhancements. Furthermore, such sensors are not compatible for reutilization. Gold nanorod arrays have recently been employed for microorganism detection, leading to moderate enhancements owing to non-resonant nature of such structures in MIR [17].

In this context, we explored the potential of Au-coated micro-structured inverted pyramid array of silicon (Au-SiIP) for the development of an efficient biosensor. Micro-structured SiIP array can be synthesized by either the photolithography or metal-assisted chemical etching (MACE) method [18-20]. The photolithography although being highly controllable and capable of patterning periodic structures, lacks in cost-effectiveness for large-scale fabrication. Alternatively, the MACE method fabricates an irregular inverted pyramid array, and it is comparably cost-efficient even for wafer-scale fabrication. It has been demonstrated that the structural irregularity can introduce broadening in the resonance spectrum and this large-wavelength domain can be utilized for the detection in a broad range. The SiIP array structures are widely used for solar cell applications and have been proven as a remarkable light-trapping structure [20]. Furthermore, surface-enhanced Raman spectroscopy (SERS) analogous to SEIRA, is recently reported the E. coli detection with $1.8 \times 10^6$ cells/ µl cell concentration employing the Au-IPs (also known as klarite) with size ranges from 1 – 1.5 µm [21]. The field confinement in IP array structures are the interplay of wavelength and IPs size. Consequently, the small dimensions of IP < 2 µm cannot effectively be resonant at MIR with wavelength range, $\lambda$ = 2.5 µm – 25 µm. Moreover, to best of our knowledge no such study is reported for bacterial detection using micro sized inverted pyramids as well for microorganism sensing exploiting MIR resonant structures within the full MIR range.

The present study is the first comprehensive exploration of relatively larger sized IP (4 µm – 8 µm) array, fabricated using the MACE method, for the detection of E. coli and S. aureus by utilizing surface-enhanced infrared absorption spectroscopy. The achieved irregularity of the SiIP array structure contributes to achieving the broad resonant spectrum which is further confirmed by the electromagnetic simulations using commercially available software, CST Microwave Studio [22]. Additionally, we presented the modification in the average intensity of the confined electric field by the change in refractive index in surrounding of SiIP array owed to different bacteria. The interpretations from these simulations are utilize for the detection of E. coli and S. aureus bacteria, each with concentrations of $4\times10^5$ cells/ µl and $8\times10^5$ cells/ µl. The highest detected concentration in our study is still lower than the detection limits established by AgNPs for SEIRA [15] and Au-IP for SERS [21]. Moreover, the detection of both is significantly important for the healthcare as both E. coli and S. aureus are common invasive infections in humans. S. aureus is gram-positive whereas E. coli is gram-negative bacteria. We quantitatively and qualitatively obtained the enhanced bacterial signals within the entire range of MIR spectrum. Additionally, the demonstrated MIR resonant structure exhibits the property of reusability and reproducibility unlike traditional microorganism sensors such as metal nanoparticles. Consequently, the proposed device illustrates adequate sensitivity and immense potential for the integration of such biosensors for future healthcare applications.

## 2. Experimental section

### 2.1. Reagent and materials

Boron-doped (1-10 Ω cm), 100-oriented c-Si wafers were purchased from Pure Wafer, USA. Ammonia solution ($NH_4OH$) and Ethanol ($C_2H_5OH$) were obtained from Sigma Aldrich. Sulphuric acid ($H_2SO_4$), Nitric Acid ($HNO_3$), Hydrofluoric acid (HF) and Hydrogen peroxide ($H_2O_2$), were bought from Thomas Baker. $Cu(SO_4)$ were purchased from Thomas Baker.

### 2.2. Experimental Section

Silicon inverted pyramid arrays were fabricated on c-Si wafers. Prior to the etching chemical reaction, the wafers were thoroughly cleaned using a three-step process to remove any organic contaminants, first being RCA ($NH_4OH$: $H_2O_2$: deionized water, DI = 1:1:5) for 10 min at 80°C, which is followed by Piranha solution cleaning ($H_2SO_4$:$H_2O_2$ = 3:1) for 10 min, consequently a thin native oxide layer formed on silicon. Therefore, in the final cleaning step, the Si wafer is immersed in 2% HF solution for 2 – 3 min. These abovementioned steps are followed by rinsing the Si wafer for at least 2-3 times with DI water. Post-cleaning for the preparation stage, wafers were immersed in a solution containing 2 mg of NaOH dissolved in 10 ml DI water for 10 min. Following this for the main etching reaction, the wafers were dipped in the etching solution for 5-8 min at 45°C in a polytetrafluoroethylene container filled with $Cu(SO_4)$:HF:$H_2O_2$:DI = 1.53:0.88:0.045g:7.7. Moving on, the residual Cu-NPs were drawn out using the nitric acid solution for 5 min. After ensuring the complete removal of Cu-NPs, the Si wafers were washed with deionized water and dried under flowing nitrogen. The SiIP array formation is done by exploiting mechanism of two half-cell reactions, cathodic reduction with $H_2O_2$ and anodic oxidation on silicon surface under Cu-NPs [17], as presented schematically in Figure 1a. The corresponding scanning electron micrograph (SEM) images of the fabricated SiIP array covering a large area along with its magnified image and cross-sectional view, are presented in Fig. 1b - 1c, respectively. The SiIP size distribution is measured by plotting histogram along with Gaussian curve-fitting from Figure 1b, gives IP average size around 5 µm. After the SiIP array fabrication, SiIP array are sputter coated with gold of thickness ~ 40 nm in order to completely exploit the light trapping phenomenon. The methodology is further discussed in the results and discussion segment.

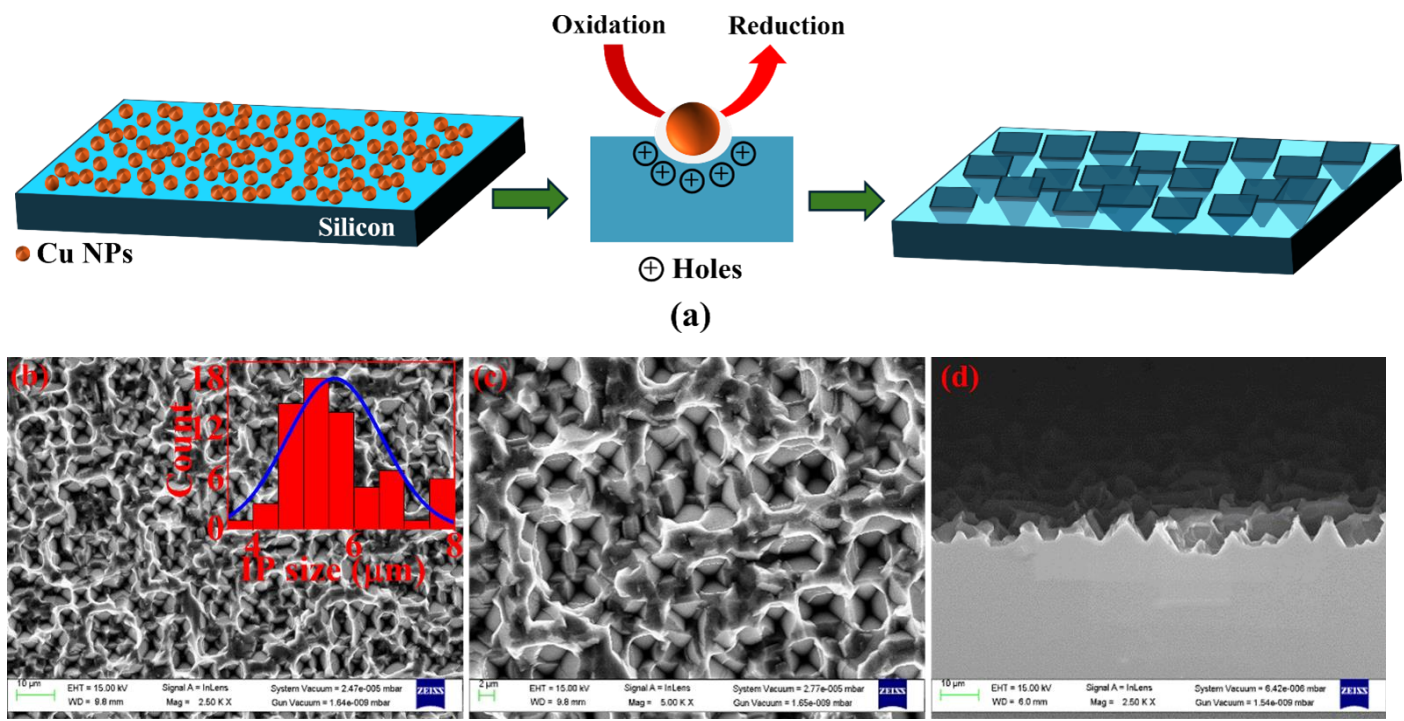

**Figure 1.** (a) Schematical illustration of the mechanism involved in the formation of Si-IP array by metal-assisted chemical etching. Scanning electron microscope image of (b, c) fabricated inverted pyramid array surface with 10 µm and 2 µm scale, respectively. The histogram shows the average size of IP ranging from ~ 3 µm – 8 µm, inset of Figure 1b. (d) the cross-sectional view of the IP array structure shown in (a) and (b).

*2.3 Bacterial preparation*

Wild type E. coli and S. aureus of strain ATCC 29213, is used for the label-free detection. The primary culture was cultivated in sterile Luria Bertani (LB) medium for both the bacteria cells, under conditions of 180 rpm and 37 °C for approximately 14 hours. Following this cultivation period, 50 µL of the solution was taken and aseptically transferred to 7 ml LB broth tubes. This transferred tubes then incubated at 180 rpm and 37 °C until an optical density of 1 is achieved. Around 20 µl bacterial solution is used for a single measurement. For the lower OD values, the appropriate dilutions are made.

*2.4. Characterization*

Fourier transform IR spectrometer (Agilent Cary 630 FTIR) was employed to perform the ATR- SEIRA measurements within the spectral range of 4000 cm$^{-1}$ to 400 cm$^{-1}$ using Diamond ATR-1-Bounce crystal. The resolution of the measurements was 4 cm$^{-1}$, incorporating 8 sample scans. The label-free detection is achieved by dropping-off as-prepared bacterial species containing 20 µL solution on the Au-SiIP. The spectra were measured at various positions on the sample. The scanning electron microscope image of the inverted pyramid array surface was taken using a Carl-Zeiss Ultra plus field emission with an accelerating voltage of 15 kV.

## 3. Results and discussion

The fabricated SiIP arrays as presented in Fig. 1b-d, are sputter coated with gold (Au) as Au acquires negative permittivity in the MIR wavelength spectral zone, facilitating the formation of surface plasmon polaritons (SPPs) on the Au/Air interface. These SPPs assists in the significant light-trapping features in Au-coated SiIP (Au-SiIP) array.

To demonstrate the light trapping capabilities of our structure, we simulated the irregular Au-SiIP array structure exploiting finite element method (FEM) based CST microwave studio software for the electro-magnetic simulations [22]. The cross-sectional area for the simulation is designed to closely resemble the cross-sectional view as depicted in Fig. 1c. We incorporated a 1 µm vacuum layer above our structure that could be the thickness of E. coli monolayer, however for our device size range the bacteria can also reside within the inverted pyramid micro-sized cavity. The performed simulations illustrate that the electric field enhancement in the inverted pyramid array covers the entire MIR spectrum region (ν =400 cm$^{-1}$ – 4000 cm$^{-1}$; λ =25 µm – 2.5 µm) as presented in Fig. 2 (The scale bar of the simulation structure is also shown in the ν =400 cm$^{-1}$ panel). The simulation is depicted for six wavenumbers (ν) within the MIR range. The electric field intensity of the confined field is the result of the interplay between inverted pyramid size and the wavelength. For λ less than or comparable to IP size, the field is supposed to be confined within the structure. However, the simulation is depicting that with increasing wavelength, the field intensity is confining to the top side or wider side of the structure. Additionally, for larger wavelengths the field in the single inverted pyramid is coupling with the field of the consecutive IP or within the overlapping IP, likely attributable to irregular structure. This is observable in ν = 666.7 cm$^{-1}$ (λ =15 µm), where the intensity is comparatively larger (yellow region) or in the additional top layer. These involved inter-IP could be crucial in explaining the broader wavelength region enhancements. Moreover, the significant field enhancements for the whole MIR spectrum are observed.

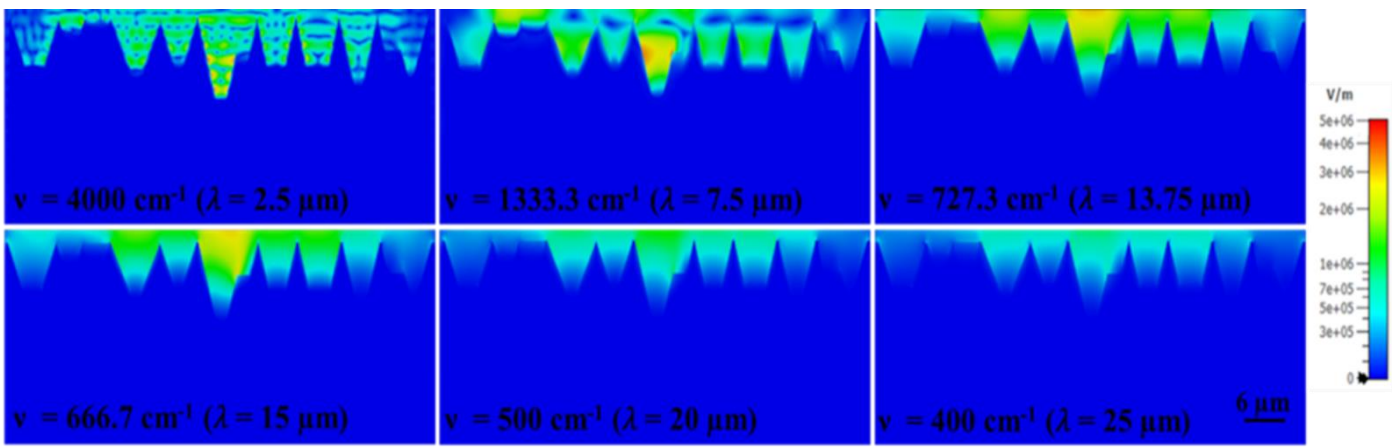

**Figure 2.** Simulated electric field enhancement for the cross-section of an inverted pyramid array for some values within the MIR range is shown. The corresponding field distribution of the electric field intensity is shown on the right side. The scale bar of 6 µm is shown in the last simulation of lower panel.

In order to demonstrate the experimental ability of these MIR resonant Au-SiIP array structures, the label-free detection of two individual bacteria, one gram-positive, E. coli and another, gram-negative, S. aureus bacteria is observed. We measured FTIR spectrum for gold coated planer silicon (Au-Si) as well as micro structured IPs silicon (Au-SiIP), with and without bacteria as presented in Fig. 3a and 3b. We measured the bacterial spectrum at two randomly chosen locations: L$_1$ and L$_2$. The spectra are measured for bacterial cell concentration of 8×10$^5$ cells/ µl having optical density, OD = 1, evaluated at λ = 600 nm. The vertical axis is plotted by subtracting the reflectance data from 100 i.e. (100-R) or the absorption as the transmission is expected to be negligible within the 40 nm Au thickness structure. The absence of any significant spectral signals for Au-Si and Au-SiIP without bacteria is observed. However, both spectra exhibit few small humps. The features observed in Au-Si are mostly suppressed in Au-SiIP, probably attributes to the presence of Au-coating. The newly identified features in Au-SiIP are observed around 2665.1, 2877.5 and 2959.5, respectively. The gold coated silicon sample with bacteria i.e., EC-Au-Si and SA-Au-Si where EC and SA are stands for E. coli and S. aureus respectively, exhibits the clear bacterial peaks compared to Au-Si and Au-SiIP. Besides this, EC-Au-SiIP and SA-Au-SiIP indicates substantial enhancement of the bacterial signals accompanied by prominent and distinct peaks. The humps observed in Au-SiIP are replicated even in bacterial sample. Moreover, the average enhancement for EC is achieved within three-fold to seven-fold range, while for SA the enhancement is two-fold to five-fold. The observed spectrum at two entirely different locations in the sample demonstrating the reproducibility of our structure. Furthermore, for Au-Si there shouldn't be any significant confinement, the observed low intensity peaks may refer to a minimal enhancement owing to the surface roughness on the gold surface.

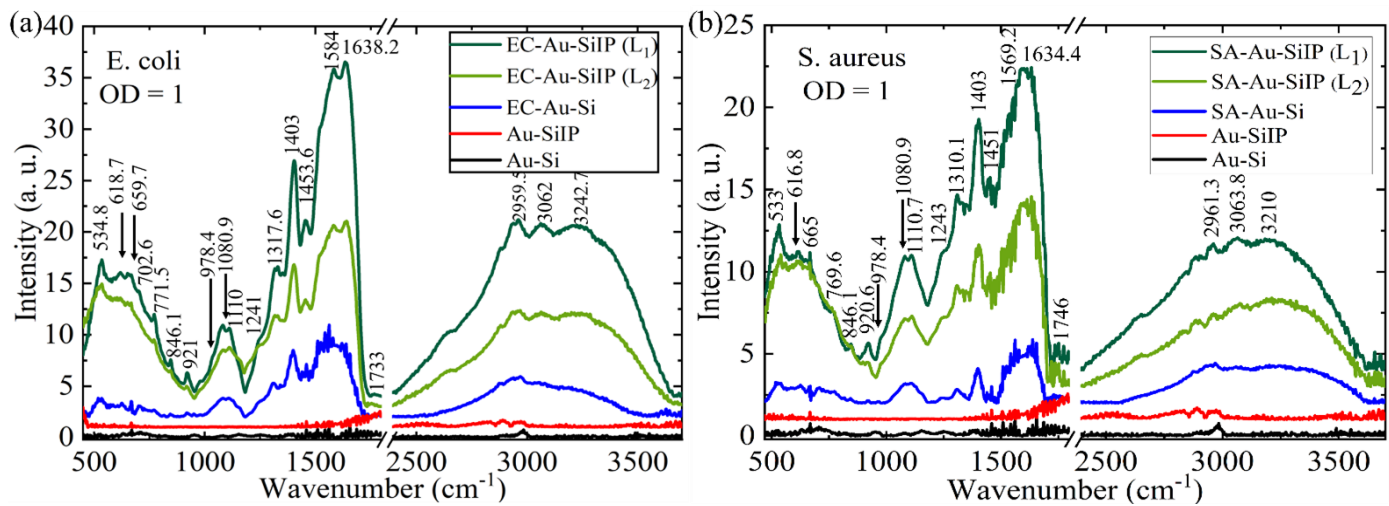

**Figure 3:** Measured FTIR spectrum for (a) E. coli (EC) and (b) S. aureus (SA) for OD =1 on Au-coated IP array (Au-SiIP) and Au-coated planer silicon (Au-Si) along with the spectrum of without bacteria. $L_1$ and $L_2$ in Figure (a) and (b) refers to two randomly chosen different locations on the sample. The corresponding peak position for each spectrum is marked. The resolution of the measurement is 4 cm$^{-1}$. All data sets are shifted by a value of 1 for more clarity.

The recorded ATR-FTIR spectrum of each bacteria represents the signals from its various bio-macromolecules, encompassing carbohydrates, lipids, DNA, and proteins. In general, in the FTIR spectrum of biological specimens, the 3000–2800 cm$^{-1}$, 1700–1500 cm$^{-1}$, 1500–1185 cm$^{-1}$, and 1185–900 cm$^{-1}$ represent distinctive spectral patterns of lipid, protein, phospholipid/DNA/RNA, and polysaccharide bands, respectively and 900–600 cm$^{-1}$ segment denotes the fingerprint region.

From our structure, for E. coli, we observed peaks at 534.8 cm$^{-1}$, 618.7 cm$^{-1}$, 659.7 cm$^{-1}$, 702.6 cm$^{-1}$ (Low), 771.5 cm$^{-1}$, 846.1 cm$^{-1}$, 921 cm$^{-1}$, 978.4 cm$^{-1}$ (Low), 1080.9 cm$^{-1}$, 1110 cm$^{-1}$, 1241 cm$^{-1}$, 1351 cm$^{-1}$, 1403 cm$^{-1}$, 1453.6 cm$^{-1}$, 1584 cm$^{-1}$, 1638.2 cm$^{-1}$, 1733 cm$^{-1}$, 2961.3 cm$^{-1}$, 3062 cm$^{-1}$ and 3227.8 cm$^{-1}$, as marked in Figure 3a. Further, for S. aureus, we observed peaks at 533 cm$^{-1}$, 616.8 cm$^{-1}$, 665 cm$^{-1}$, 769.6 cm$^{-1}$, 846.1 cm$^{-1}$, 920.6 cm$^{-1}$, 978.4 cm$^{-1}$ (Low), 1080.9 cm$^{-1}$, 1110.7 cm$^{-1}$, 1243 cm$^{-1}$, 1310.1 cm$^{-1}$, 1403 cm$^{-1}$, 1451 cm$^{-1}$, 1518 cm$^{-1}$ (Low), 1634.4 cm$^{-1}$, 1746 cm$^{-1}$, 2961.3 cm$^{-1}$, 3063.8 cm$^{-1}$ and 3209.2 cm$^{-1}$, as marked in Figure 3b. Some of the peaks have low intensity, written low in brackets next to them. One peak at ~ 702.6 cm$^{-1}$ is missing in S. aureus. The band-assignment of the identified peaks is presented in supplementary table 1.

Furthermore, few EC's peak positions do not align with those of SA, nonetheless the observed shift does not follow a systematic pattern. It may be attributed to the unique fundamental characteristics of individual bacteria. Furthermore, the enhanced signal intensity observed in E. coli spectrum as compared to S. aureus, could be ascribed to three distinct factors: (i) Different location of the conducted measurement, (ii) unique dielectric constants of bacteria as the field confinement is contingent upon the dielectric properties of metal as well the surrounding medium, (iii) Characteristic structural configuration. However, the IP cavity size is analogous to bacteria size, it is anticipated that (ii) and (iii) should exhibit similar effects on the measurement. We investigated this aspect and present the corresponding simulations in supplementary data. The dielectric permittivity for E. coli and S. aureus is, $\varepsilon \approx$ 2.86 and $\varepsilon \approx$ 4.11, respectively [23]. The simulations are performed on Au-SiIP array by considering equivalent bacterial dielectric properties. The associated confine field is correlated for wavenumbers ν = 4000 cm$^{-1}$, ν = 666.7 cm$^{-1}$ and ν = 400 cm$^{-1}$, are presented in Figure S1. The maximum field computed for E. coli in the x-y plane for ν = 4000 cm$^{-1}$, ν = 666.7 cm$^{-1}$ and ν = 400 cm$^{-1}$ is 4.571×10$^6$ V/m, 4.553×10$^6$ V/m and 2.9 ×10$^6$ V/m respectively, whereas for S. aureus, it is 3.804×10$^6$ V/m, 4.805×10$^6$ V/m and 2.86×10$^6$ V/m respectively. We noticed that until ν = 727.3 cm$^{-1}$, the field strength is comparatively elevated for the lowest dielectric constant (Figure 2 and S1 (a-b)), following this up to ν = 666.7 cm$^{-1}$, the field is now uplifted for ε = 2.86 and then the higher field is achieved for ε = 4.11. It is likely that the higher dielectric permittivity facilitating the expansion of the electric field throughout and within the structure, as observed by the confinement regions, in the vicinity of the interface for the ε = 1, Figure 2.

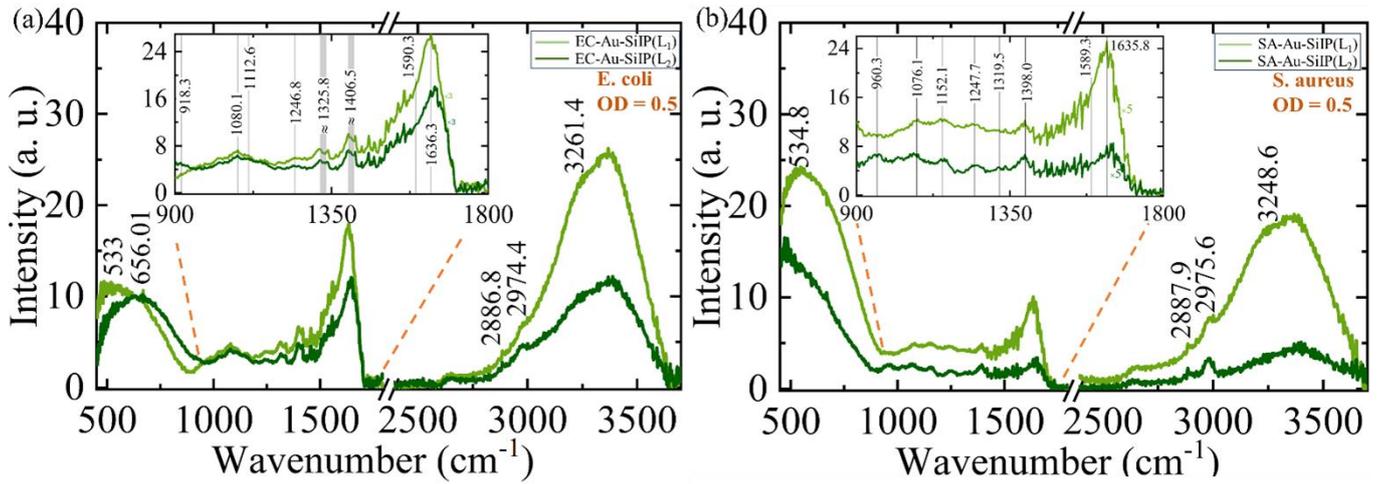

**Figure 4.** Measured FTIR spectrum for OD = 0.5, (a) for E. coli and, (b) for S. aureus. $L_1$ and $L_2$ refers to two randomly chosen different locations on the sample. The resolution of the measurement is 4 cm$^{-1}$.

However, the observed change is not substantial, as the metal is dominantly responsible for localization of the electric field. Consequently, given the small change in magnitude, it is reasonable to anticipate the insignificant impact on the signal intensity. Additionally, the cross-sectional field confinement view in Figure 2 and S1 (a-b), features the structure-dependent confinement which can significantly alter the signal intensity. Accordingly, reason (i) is more aligned along with the reenforcing measurements and simulations. Following these measurements, we measured the FTIR spectrum for the lower bacterial cell concentration of 4×10$^5$ cells/ µl (0.5 OD) and the corresponding FTIR spectrum is presented in Figure 4. The signals detected are still providing almost all information throughout spectrum. The signal intensity is well-matched with the spectrum of OD =1 except in the region from 900 – 1500 cm$^{-1}$. In the fingerprint region from 500 – 900 cm$^{-1}$, the features are merged into a broad signal likely due to low intensities. However, from 900 – 1800 cm$^{-1}$, almost all the features convey detailed information as presented in the insets of Figure 4. Furthermore, the spectrum from 2500 – 3500 cm$^{-1}$, is also significantly replicated the peaks with the exception of overlapping peak around 3069.1 cm$^{-1}$. Each identified peak is corresponding to the bacterial FTIR assessment discussed previously.

## Conclusion

We investigated the gold-coated micro-structured inverted pyramid array of silicon, aiming to leverage their potential light-trapping capabilities for the enhanced light-matter interaction by employing surface plasmon polariton in the mid-infrared range. Concretely, we label-free achieved the quantitative and qualitative detection E. coli and S. aureus bacteria. The detection of both the bacterial species is achieved for the cell concentration of 4×10$^5$ cells/ µl (OD = 0.5) and 8×10$^5$ cells/ µl (OD = 1). Along with the comprehensive observed enhanced spectrum, the key attribute involves the distinct amplified signals in fingerprint region. Furthermore, our MIR resonant structure has the characteristic capability of reusability and reproducibility unlike conventional microorganism sensors. The explored biosensors are well aligned with standard CMOS technology, underscoring their capabilities in the advancement of forthcoming bio-sensing technologies.

### Conflicts of interest

There are no conflicts to declare.

### Acknowledgement

P. Sudha, A. Kumar and K. Ansari acknowledges Council of Scientific and Industrial Research (CSIR) for financial support during this research. K. Dhankar acknowledges Ministry of Human Resource and Development (MHRD) for financial support during this research.

# Large-Scale Cost-Effective Mid-Infrared Resonant Silicon Microstructures for Surface-Enhanced Infrared Absorption Spectroscopy

Pooja Sudha[1], Anil kumar[1], Kunal Dhankar[2], Khalid Ansari[1], Sugata Hazra[2,3], Arup Samanta[1,3]

[1]Department of Physics, Indian Institute of Technology Roorkee, Roorkee, Uttarakhand 247667, India
[2]Department of Biotechnology, Indian Institute of Technology Roorkee, Roorkee, Uttarakhand 247667, India
[3]Centre for Nanotechnology, Indian Institute of Technology Roorkee, Roorkee, Uttarakhand 247667, India.

**Simulated electric-field enhancement in the SiIP-sensor for the bacterial surrounding:**

We also simulated the SiIP array by changing the surrounding dielectric constant around our structure, which has taken as identical to bacteria' dielectric properties. The simulated results are shown in the Figure S1. These results are discussed in the main text in detail.

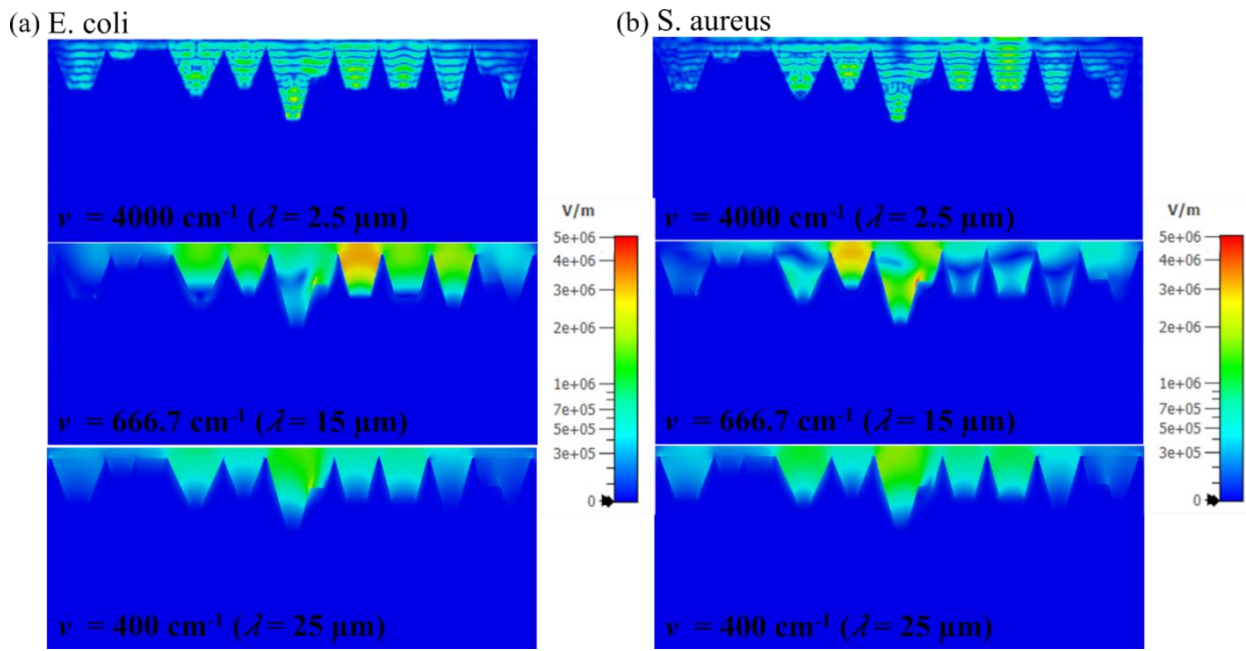

**Figure S1.** Simulated electric field enhancement for the surrounding dielectric properties that of bacteria around the SiIP array structure (a) for E. coli and (b) for S. aureus.

**Band assignment corresponding to observed SEIRA peaks:**

Table S1 represents the microorganism band assignment corresponding to observed SEIRA peaks [1, 2].

**Table S1.** The band-assignment of SEIRA peaks

| FTIR band in wavenumber (cm$^{-1}$) | E. coli ($8 \times 10^5$ cells/μL) | S. aureus ($8 \times 10^5$ cells/ μL) |
|---|---|---|
| Carbohydrates | 534.8 | 533.01 |
| Aromatic ring skeletal | 618.7 | 616.8 |
| S-S bond (membrane proteins) | 659.7 | 665 |
| Guanine | 702.6 | 702.6 |
| C-H bending: nucleic acids | 771.5 | 769.6 |
| L-Tryptophan | 846.1 | 846.1 |
|  | 921 | 920.6 |
| C-N$^+$-c stretch: nucleic acids | 978.4 | 978.4 |
| PO$^{-2}$ symmetric: nucleic acids and phospholipids | 1080.9 | 1080.9 |
| Adenine | 1110 | 1110.7 |
| P=O stretching | 1241 | 1243 |
| | 1317.6 | 1310.1 |
| COO- symmetric stretch | 1403.3 | 1403 |
| $CH_2$ bending: lipids | 1453.6 | 1451 |
| | 1520.7 | 1517.0 |
| Amide II(protein N-H bending, C-N stretching): α-helices | 1584 | 1569.2 |
| Amide I (protein C=O stretching): α-helices | 1638.2 | 1634.4 |
| Ester C=O stretch: phospholipids, triglycerides | 1733 | 1746 |
| $CH_2$ symmetric stretch: mainly lipids | 2879 | 2884.95 |
| $CH_3$ asymmetric stretch: mainly lipids | 2959.5 | 2961.3 |
| C-H stretching | 3062 | 3063.8 |
| N-H and O-H stretching vibration: proteins, polysaccharides | 3242.7 | 3210 |